\documentclass[reprint,superscriptaddress,amsmath,amssymb,aps,pra]{revtex4-2}

\usepackage{graphicx}% Include figure files
\usepackage{dcolumn}% Align table columns on decimal point
\usepackage{bm}% bold math

\usepackage{graphicx}% Include figure files
\usepackage{epstopdf}
\usepackage{dcolumn}% Align table columns on decimal point
\usepackage{bm}% bold math
\usepackage{braket}
\usepackage{color}
\usepackage{mathrsfs}
\usepackage{ragged2e}
\usepackage{eqnarray}
\usepackage{subeqnarray}
\usepackage{cases}
\usepackage{amsfonts}
\usepackage{dcolumn}
\usepackage{bm}
\usepackage{tikz}
\usepackage{float}
\usepackage[colorlinks=true,citecolor=blue,urlcolor=blue]{hyperref}
\usepackage{amsmath,amsfonts,amssymb,times,natbib}
\usepackage[standard]{ntheorem}
\usepackage{natbib}

\newcommand*{\red}{\textcolor{red}}

\begin{document}

\preprint{APS/123-QED}

\title{All-optical convolution utilizing processing in memory based on a cold atomic ensemble}

\author{Ying-Hao Ye}
\affiliation{Anhui Province Key Laboratory of Quantum Network, University of Science and Technology of China, Hefei 230026, China.}%
\affiliation{Laboratory of Quantum Information, University of Science and Technology of China, Hefei 230026, China.}%
\affiliation{Hefei National Laboratory, University of Science and Technology of China, Hefei 230088, China.}%

\author{Jia-Qi Jiang}
\affiliation{Wang Da-Heng Center, Heilongjiang Key Laboratory of Quantum Control, Harbin University of Science and Technology, Harbin 150080, China.}%

\author{En-Ze Li}
\affiliation{Anhui Province Key Laboratory of Quantum Network, University of Science and Technology of China, Hefei 230026, China.}
\affiliation{Laboratory of Quantum Information, University of Science and Technology of China, Hefei 230026, China.}
\affiliation{Hefei National Laboratory, University of Science and Technology of China, Hefei 230088, China.}

\author{Wei Zhang}
\affiliation{Anhui Province Key Laboratory of Quantum Network, University of Science and Technology of China, Hefei 230026, China.}%
\affiliation{Laboratory of Quantum Information, University of Science and Technology of China, Hefei 230026, China.}%
\affiliation{Hefei National Laboratory, University of Science and Technology of China, Hefei 230088, China.}%

\author{Da-Chuang Li}
\affiliation{Institute for Quantum Control and Quantum Information and School of Physics and Materials Engineering, Hefei Normal University, Hefei, Anhui 230601, China}%

\author{Zhi-Han Zhu}
\email{zhuzhihan@hrbust.edu.cn}
\affiliation{Wang Da-Heng Center, Heilongjiang Key Laboratory of Quantum Control, Harbin University of Science and Technology, Harbin 150080, China.}%

\author{Dong-Sheng Ding}
\email{dds@ustc.edu.cn}
\affiliation{Anhui Province Key Laboratory of Quantum Network, University of Science and Technology of China, Hefei 230026, China.}%
\affiliation{Laboratory of Quantum Information, University of Science and Technology of China, Hefei 230026, China.}%
\affiliation{Hefei National Laboratory, University of Science and Technology of China, Hefei 230088, China.}%

\author{Bao-Sen Shi}
\email{drshi@ustc.edu.cn}
\affiliation{Anhui Province Key Laboratory of Quantum Network, University of Science and Technology of China, Hefei 230026, China.}%
\affiliation{Laboratory of Quantum Information, University of Science and Technology of China, Hefei 230026, China.}%
\affiliation{Hefei National Laboratory, University of Science and Technology of China, Hefei 230088, China.}%

\date{\today}% It is always \today, today,

\begin{abstract}
Processing in memory (PIM) has received significant attention due to its high efficiency, low latency, and parallelism. In optical computation, coherent memory is a crucial infrastructure for PIM frameworks. This study presents an all-optical convolution experiment conducted within computational storage based on a cold atomic ensemble. By exploiting the light-atom phase transfer facilitated by the electromagnetically induced transparency, we demonstrated spiral phase contrast processing of photon images in memory, resulting in the edge enhancement of retrieved images recorded using time-correlated photon imaging. In particular, adopting state-of-the-art atomic techniques provides a coherent memory lifetime exceeding 320 µs for PIM operations. Our results highlight the significant potential of cold atomic ensembles as computational storage for developing all-optical PIM systems.
%%Processing in memory (PIM) has garnered significant attention due to its high efficiency, low latency, and advantages in parallel processing. In the context of all-optical PIM, atomic ensembles are excellent candidates for storing photonic information because of their long coherence times in hyperfine ground levels and their flexibility in manipulation through light fields. 
%%In this study, we adopt the well-established electromagnetically induced transparency protocol and propose a scheme for implementing PIM using atomic media. We demonstrate our proposal using an edge enhancement technique called spiral phase contrast (SPC). Our setup involves an intensified charge-coupled device that captures the read-out signal field, benefiting from SPC's high sensitivity to phase or intensity gradients and its interferometric nature, which allows for detection without compromising brightness.
%%After mapping the photonic information to magnetically insensitive states and applying a guiding magnetic field to counteract the nonuniform evolution of collective excitation, we find that the retrieved signal pattern remains recognizable even after a storage duration of over 320 $\mu s$.
\end{abstract}

%\keywords{Suggested keywords}%Use show keys class option if keyword
\maketitle

%\tableofcontents

\section{\label{sec:level1}Introduction}
Cold atomic ensembles present promising light-matter interfaces for all-optical processing, primarily due to their collective enhancement effects \cite{collectiveeffect}, consistent spectral properties, and reduced Doppler broadening. The enhanced strength of light-matter interactions simplifies the otherwise stringent experimental requirements, such as high-Q resonators, for increasing the absorption cross-section of a single-atom system \cite{kimble1998strong} and facilitates operations in specific optical modes \cite{ZHAOcollectivelyexcitedatomic}. Moreover, the consistent spectral characteristics, combined with the suppression of Doppler broadening, offer significant advantages for constructing and scaling atomic-based networks \cite{quantumrepeaterreview}.

One of the most compelling applications of cold atomic ensembles is their use as storage media for photonic information \cite{CHANELIERE201877,opticalquantummemoryNP}. The long coherence times of atomic metastable states and the reduced particle drift in low-temperature systems support robust optical storage \cite{longlivedmemory}, provided that ambient magnetic fields are precisely controlled \cite{Dsprevivalexperiment,Dsprevivaltheory,Dsprevivalexperiment}. 
Researches have demonstrated long-lived and faithful memory across various photonic degrees of freedom, including polarizations \cite{highefficiencypolarizationmemory}, wavevector directions \cite{wavevectordirectionmemory}, and temporal modes \cite{temporalmodequantummemory}. Among these, the transverse modes of the stored field are particularly promising for image-correlation applications, offering tremendous flexibility for encoding information \cite{OAMencodingterabit}. To date, studies have reported successful storage of complex intensity \cite{Imagestoreinhotatom} or phase patterns of photons and have even extended into the realm of single-photon storage \cite{OAMqubitstorage,Ding2013single}.

A widely used protocol for memorizing transverse multimode data employs the optical modulation technique known as electromagnetically induced transparency (EIT) \cite{EITRev}. In this approach, photonic information is actively transferred to the profile of metastable collective atomic excitations, or spin waves, in a manner that is lossless and reversible \cite{collectiveeffect}. The processes of storage (write-in) and retrieval (read-out) can be mathematically described using Heisenberg equations \cite{EITDSPPRL,EITDSPPRA}. By incorporating transverse spatial degrees of freedom into these equations, we can identify methods to manipulate the collective excitations’ profile by tailoring the wavefronts of the control fields. 
Cold atom ensembles provide essential infrastructure for all-optical processing within Processing-in-Memory (PIM) frameworks due to their storage capacity and potential for flexible manipulation. As a burgeoning area of research, PIM systems offer significant advantages, such as breaking the von Neumann bottleneck \cite{breakvonneumannbottleneck}, improved efficiency, and parallelism by alleviating the latency and energy cost due to data transmission between processing and memory units.

\begin{figure*}[t]
	\centering
	\fbox{\includegraphics[width=\linewidth]{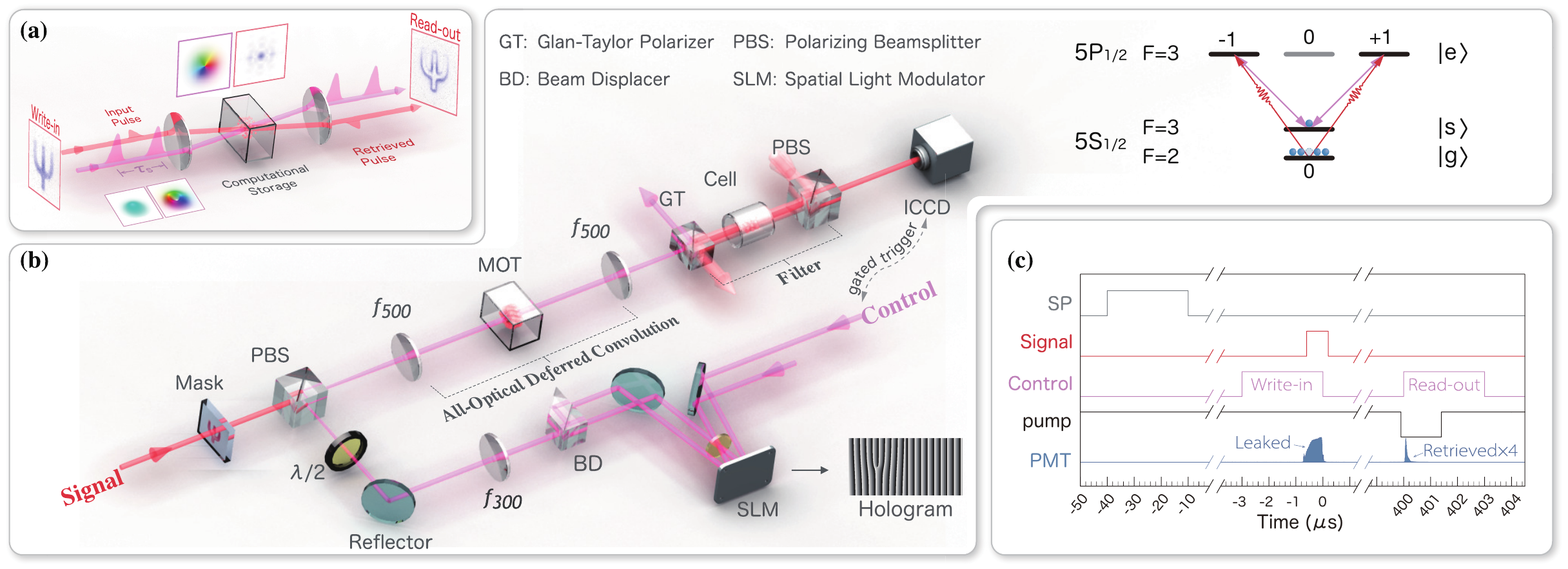}}
	\caption{Experimental scheme. (a) Conceptual graph of all-optical convolution in computational storage. (b) Schematic representation of the experimental setup, highlighting key components such as the magneto-optical trap (MOT), image target (Mask), spatial light modulator (SLM), and an intensified charge-coupled device (ICCD) operating in the Time Correlated Single Photon (TCSP) imaging mode. The energy level structure of the EIT protocol is depicted in the upper right corner. (b) An example of time sequence adopted for a storage time of 400 $\mu s$, together shown the signal pulse registered by a PMT.}
	\label{Fig:principle}
\end{figure*}

The transfer of spiral phases between collective excitation and photons has been theoretically proposed in an off-resonant Raman memory protocol \cite{Ramanstorageoamconversion}. Here, we generalize this concept to the on-resonance EIT protocol and report an experimental demonstration for optical PIM based on an atomic medium by performing all-optical convolution of images in memory. The feasibility of our approach is illustrated through an edge-detection technique called spiral phase contrast (SPC) \cite{SPCinmicroscopy},in which the vortex phase of control fields is 'imprinted' onto spin waves during both the write-in and read-out stages, acting as modulation transfer functions on the Fourier spectrum. 
A 4$f$ imaging setup performs the convolution between the original patterns and the vortex phases, and the read-out pattern are faithfully captured by an intensified charge-coupled device (ICCD) without sacrificing brightness, thanks to the interferometric and centrally symmetric nature of the SPC process \cite{radiallyHTtransform}. For a longer attainable on-demand duration of storage, we minimize destructive interference among different Zeeman sub-levels of atomic coherence by mapping photonic information to a pair of magnetically insensitive states. Additionally, we apply a strong guiding magnetic field to suppress inhomogeneous evolution across the cross-section \cite{Synchronizedye}. We finally achieved on-demand read-out of processed images for over 320 $\mu s$, which is, to the best of our knowledge, the most extended duration reported for an atomic-based memory to date. 

\section{Principle and Methods}
FIG.~\ref{Fig:principle} (a) illustrates the concept of all-optical convolution within the PIM framework, wherein a cold atomic ensemble is positioned at the spectral (or Fourier) plane of the 4f optical convolution system. This ensemble serves as the memory medium for all-optical PIM, capable of performing convolution computations, storing results, and subsequently releasing them upon request for further processing. FIG.~\ref{Fig:principle} (b) depicts a more detailed scheme of the experimental setup, where coherence memory is implemented using the EIT protocol based on a Lambda-type level structure, as illustrated in the upper right corner of FIG.~\ref{Fig:principle} (b). The states $\left|g\right\rangle$, $\left|s\right\rangle$ and $\left|e\right\rangle$ correspond to the hyper-fine energy levels $\left|5S_{1/2}, F=2\right\rangle$, $\left|5S_{1/2}, F=3\right\rangle$ and $\left|5P_{1/2}, F=3\right\rangle$ of Rb85, respectively. The signal field ($\mathbf{E}(\vec{\mathbf{r}})$) and the control field ($\mathbf{U}(\vec{\mathbf{r}})$) resonantly address the transitions $\left|g\right\rangle \leftrightarrow \left|e\right\rangle$ and $\left|s\right\rangle \leftrightarrow \left|e\right\rangle$. The entire process of storage and retrieval is commonly represented as the propagation of quasi-bosons, known as Dark State Polaritons (DSPs), within the storage medium \cite{EITDSPPRL,EITDSPPRA}. The composition of the photonic and atomic excitation components in a DSP, along with its group velocity, is governed by the Rabi frequency of the control light \cite{EITDSPPRL,EITDSPPRA}. During the write-in stage, DSPs become purely atomic, and their group velocity drops to zero as the control light is adiabatically turned off. This transition facilitates the transfer of the spatial profile from the light fields into atomic excitation. In the subsequent read-out stage, the control light is adiabatically reintroduced, causing the DSPs to become photonic with its group velocity increases until they escape the atomic medium as retrieved light. An example of signal field registered by a photomultiplier tube (PMT) and corresponding timing sequence is shown in FIG.~\ref{Fig:principle} (c). 

We use SPC convolution as an illustrative example, demonstrate its capability for edge enhancement in our PIM setup, and clarify the principle behind it. Unlike edge detection techniques that rely on differential computation \cite{NomarskiDIC,zhouzhiyuanedgedetectionquantun}, SPC is a phase-only convolution method, which theoretically allows for greater transmission efficiency. In our scheme of all-optical PIM, the spiral phase $\mathrm{exp}(i l \phi)$ can be bi-directionally transmitted between light fields and collective atomic excitations during the write-in and read-out stages; readers may refer to the supplementary material for a theoretical framework. In summary, the spatial profile of the collective atomic excitation after the write-in stage is proportional to $\mathcal{E}_{in}^{S}(\vec{\bm{r}})\left[\mathcal{U}_{l}^{W}(\vec{\bm{r}})\right]$. Here $\mathcal{E}_{in}^{S}(\vec{\bm{r}})=\mathscr{F}\left[E(\vec{\bm{\rho}})\right]$ represents the signal field within the ensemble, which is determined by the Fourier spectrum of the input pattern. The term $\mathcal{U}_{l}^{W}(\vec{\bm{r}})=g(r)\mathrm{exp}(-i l \phi)$ denotes the control light field in the write-in stage and is characterized by a Gaussian amplitude $g(r)$ and a vortex topological charge $l$, also referred to as the hyper-geometric Gaussian (HyGG) mode \cite{Hyggmode}; here $\vec{\bm{\rho}}=\vec{\bm{\rho}}(\rho,\varphi)$ and $\vec{\bm{r}}=\vec{\bm{r}}(r,\phi)$ represents the transverse coordinates in the mask plane and the Fourier plane (coincides with the central of the ensemble), respectively. $\rho$ and $r$ denote radial coordinates, while $\varphi$ and $\phi$ denote azimuthal coordinates. For the specific case of SPC, where $\left|l\right|=1$, the resulting processing is analogous to a radial Hilbert filter \cite{radiallyHTtransform}. 

After a tunable storing period ($\tau_s$), a second control pulse featuring a spatial profile $\mathcal{U}_{l'}^{R}(\vec{\bm{r}})$ is introduced to read out the stored information in the atomic coherence. This process simultaneously imprints a spiral phase onto the retrieved signal field, with chirality opposite to that observed during the write-in stage \red{[XXX]}. The second lens of the optical convolution system ($f_{500}$ in FIG.~\ref{Fig:principle} (b) following the MOT) acts as an inverse Fourier transformer, projecting the processed image onto the imaging plane. According to the convolution theorem, the retrieved signal field captured by the camera is represented by 
\begin{equation}
	\label{eqn:retrievedfield}
	E\left(-\vec{\bm{r}}\right)\ast\mathscr{F}^{-1}\left(\mathcal{U}_{-l}^{W}\mathcal{U}_{l'}^{R}\right)\propto E\ast \mathscr{F}^{-1}\left[g^2(r)\mathrm{exp}(i \Delta l \phi)\right]
\end{equation}
where the asterisk denotes the convolution operation, and $\Delta l = \left|l-l^{\prime}\right|$ signifies the difference between the topological charges of the control light in the write-in and read-out stages. The point spread function $\mathscr{F}^{-1}\left[g^2(r)\mathrm{exp}(i \Delta l \phi)\right] \propto \mathrm{exp}(i \Delta l \phi)\mathcal{H}_{\Delta l}\left[g^2 (r)\right]$ characterizes the far-field pattern of a HyGG mode, which bears resemblance to the doughnut-shaped Laguerre-Gaussian mode. In the specific case where $\Delta l = 1$, as shown in FIG.~\ref{Fig:principle} (a), consider any arbitrary point within the uniform region of the signal field. The contributions from two points that are symmetrically located around this point cancel each other out due to destructive interference. However, near the boundary regions where intensity or phase discontinuities occur, the conditions for destructive interference do not apply. This leads to a phenomenon known as edge enhancement \cite{SPCandshg}.

The signal and control light are combined using a polarizing beam splitter (PBS) and and interact collinearly within an ensemble of cold rubidium 85 atoms confined in a magneto-optical trap (MOT), as illustrated in FIG.~\ref{Fig:principle} (b). A digital beam shaping device is utilized to program the wave fronts of the control light sequences sequentially, i.e., $\mathcal{U}_l^W (\vec{\bm{r}})$ and $\mathcal{U}_{l'}^R (\vec{\bm{r}})$. \red{This device primarily consists of a spatial light modulator (SLM) and a beam displacer (BD), which transforms a pair of parallel beams with orthogonal polarizations into a desired vector mode \cite{Zhuzhihanvectormode}.} The spiral phase from the SLM is projected onto the Fourier plane using an imaging system composed of lenses $f_{300}$ and $f_{500}$, effectively overlapping the spatial spectrum of the mask carried by the signal field. Given that the spatial dimensions of the atomic medium ($\sim$2mm in diameter) are significantly smaller than the Rayleigh length of the signal field ($\sim$100 mm), we limit the numerical analysis to the 2D Fourier plane for simplicity \red{[XXX]}. The retrieved signal field is captured using an intensified charge-coupled device (ICCD) operating in Time-Correlated Single Photon (TCSP) imaging mode. The temporal width of the retrieved signal is determined by the intensity of the control light \cite{EITretrievedpulsewidth}. We optimize the signal-to-noise ratio by matching the temporal width of the retrieved pulse (~$10^2$ ns in FWHM) to the exposure time of the ICCD. This is achieved by adjusting the Rabi frequency of control light ($\Omega_{C}\approx 4\Gamma$, where $\Gamma$ represents the natural linewidth of the D1 transition of Rb85 \cite{Rb85Dlinedata}). To filter out the intense control light while preserving the transverse structure of the relatively weak retrieved signal ($\sim$O (10 nW)), we employed a narrow bandpass filter, which consists of a Glan-Taylor polarizer and a rubidium absorption cell. A pump light, which counter-propagates with the signal light and is resonant with the $\left|5S_{1/2}, F=2\right\rangle \leftrightarrow \left|5P_{1/2}, F=2\right\rangle$ transition, continuously pumps the atoms in the absorption cell to $\left|5S_{1/2}, F=3\right\rangle$, resulting in strong absorption of the control light. The pump light is turned off during the retrieval window, as shown in FIG.~\ref{Fig:principle} (c), to prevent undesired scattering noise.

\section{Results}
\begin{figure}
	\centering
	\fbox{\includegraphics[width=\linewidth]{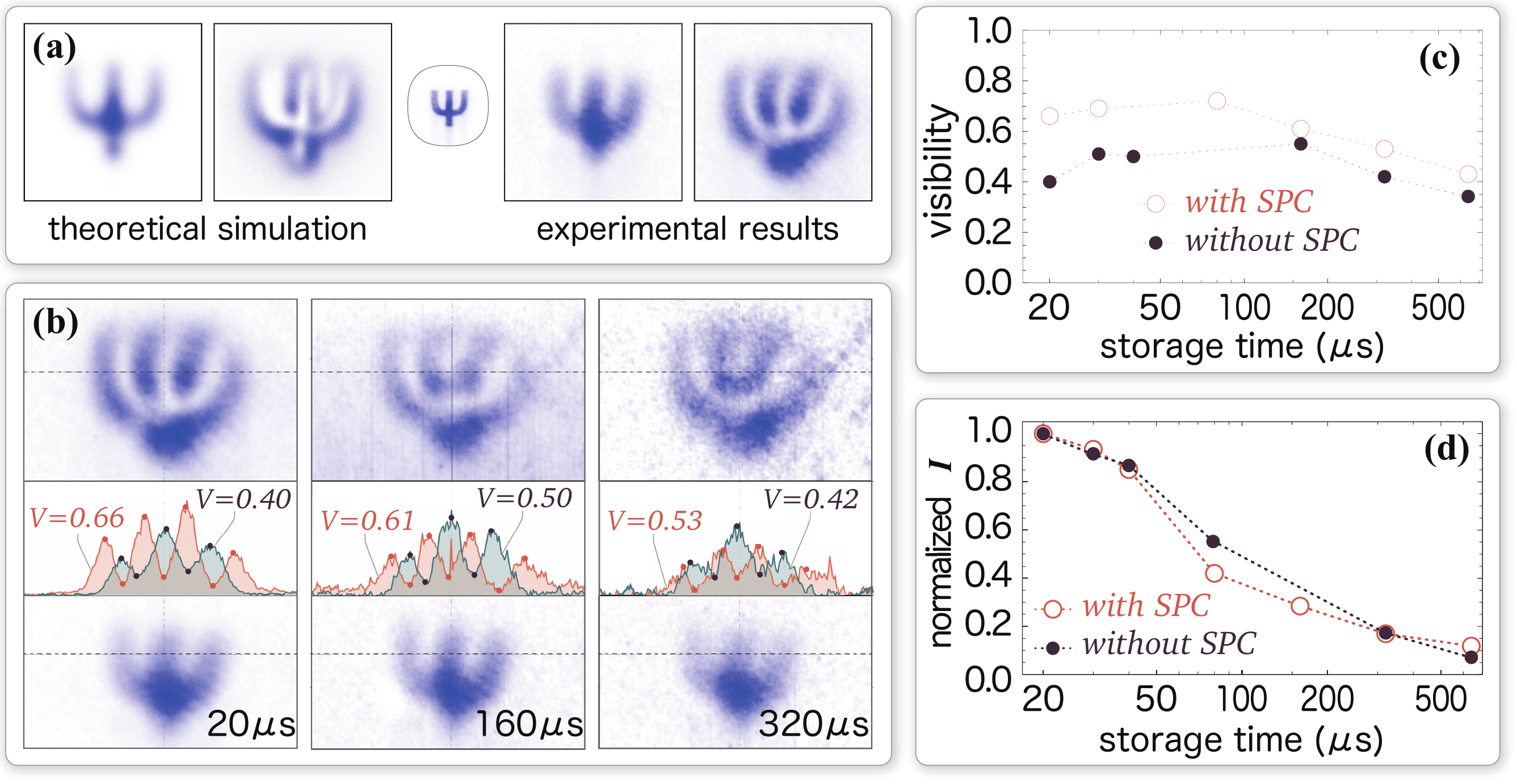}}
	\caption{Experimental results of all-optical convolutions performed during the write-in stage. Panel (a) shows both simulated and experimental convolution images retrieved from the ensemble, while panels (b) through(d) depict patterns, image visibility, and photon counts of the retrieved images after various storage durations, respectively.}
	\label{Fig:temporalcharacter}
\end{figure}

\emph{All-optical processing in the write-in Stage. — } FIG.~\ref{Fig:temporalcharacter} shows the experimental results during the write-in stage, focusing on the temporal characteristics of image convolution within our PIM setup. We choose the Greek letter ‘$\bm{\Psi}$’ as the example pattern and use two types of control light during the write-in stage: (i) a fundamental Gaussian mode ($\mathcal{U}_{l=0}^W$) for low-pass spectrum filtering, and (ii) a HyGG ($\mathcal{U}_{l=1}^W$) mode for edge enhancement processing. In FIG.~\ref{Fig:temporalcharacter} (a), we display the observed signal pattern ‘$\bm{\Psi}$’ alongside the theoretical simulation based on our experimental parameters \red{[XXX]}.The retrieved images align well with the theoretical simulations and remain recognizable even after a storage duration exceeding 300 $\mu s$, as shown in FIG.~\ref{Fig:temporalcharacter} (b), which corresponds to a transmission distance of over 60 kilometers in optical networks. To quantitatively assess the degradation of the retrieved image over time, we employed a metric called "image visibility," defined as  $V=(I_{max}-I_{min})/(I_{max}+I_{min})$. Here, $I_{max}$ and $I_{min}$ represent the average values of the local maxima and minima, respectively, measured along a sampling slice indicated by the dotted-dashed line in the retrieved images.Each extremum is the average of the first three maximum or minimum photon counts recorded by individual pixels along the sampling slice. The image visibilities for various storage times are presented in FIG.~\ref{Fig:temporalcharacter} (b), with the corresponding extremums highlighted for clarity.

The results of $V$ versus storage time are summarized in FIG.~\ref{Fig:temporalcharacter} (c),  indicating that retrieved patterns subjected to edge-enhancement processing display greater visibility than those subjected to low-pass filtering. This phenomenon can primarily be attributed to two mechanisms: the long-lived transverse coherence of collective atomic excitations and the high photon utilization efficiency inherent in the SPC-based edge enhancement, which is a phase-only convolution technique. In FIG.~\ref{Fig:temporalcharacter} (d), we illustrate the integrated photon counts of retrieved images over the storage period, confirming that the decoherence of atomic collective excitations is minimally affected by the inclusion of SPC convolution.  However, when storage time exceeds 500 $\mu s$, the retrieved image becomes nearly indistinguishable due to a decline in the signal-to-noise ratio. This degradation is mainly caused by a continuous decrease in retrieval efficiency and the presence of unwanted scattered photons from the absorption cell. To mitigate these adverse effects, a more sophisticated experimental setup may be implemented. Specifically, retrieval efficiency can be enhanced by increasing the optical depth of the storage media through techniques such as dark traps \cite{darkmotfirst,highdensitydark1993} and optical depumping, which facilitate greater particle accumulation in the desired initial state \cite{ultrahighOD}. Additionally, lowering the temperature of atoms using a technique known as "optical molasses" could help minimize particle loss during storage \cite{opticalmolassestheory,opticalmolassesexperiment}.
\begin{figure}
	\centering
	\fbox{\includegraphics[width=\linewidth]{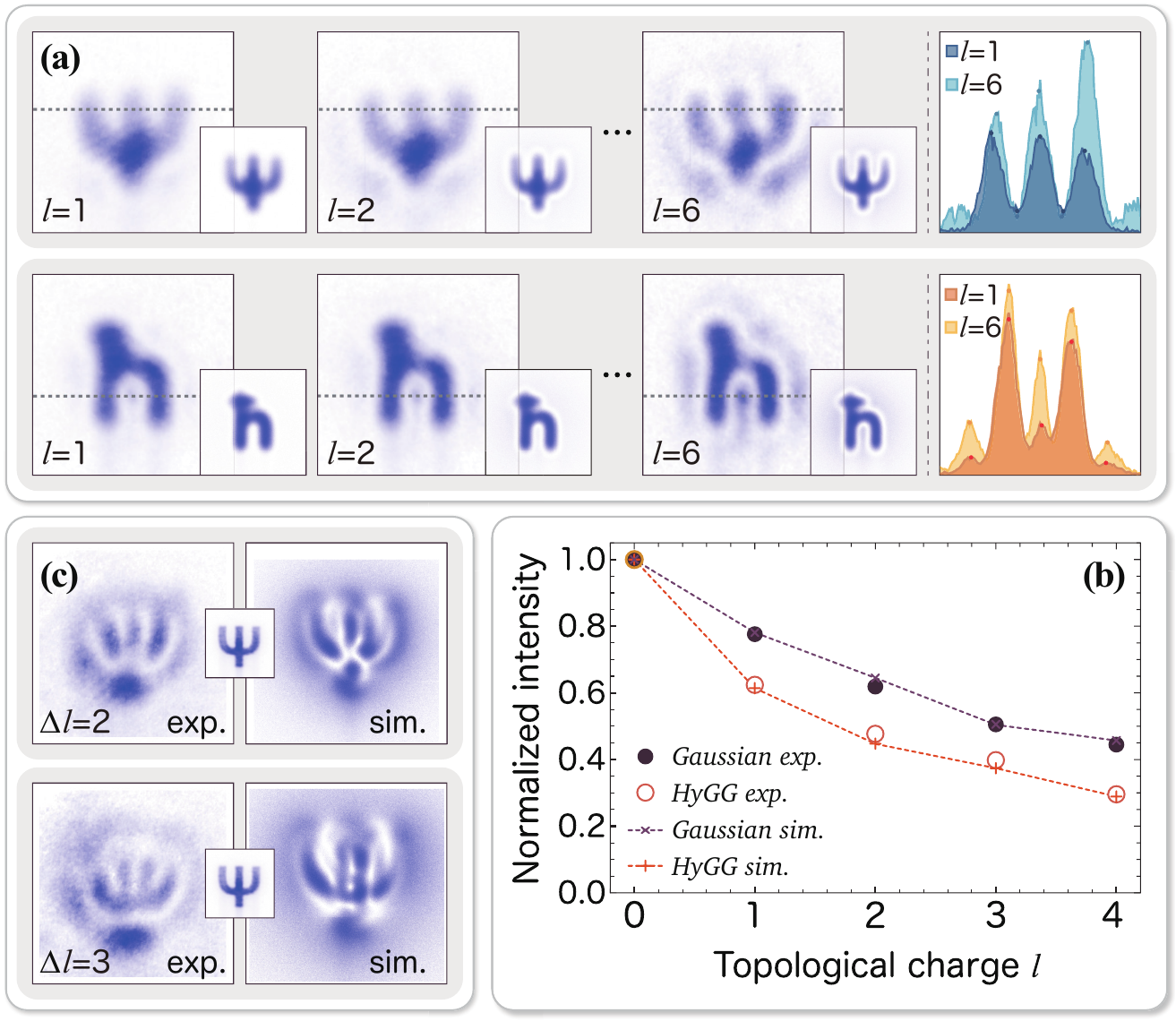}}
	\caption{Experimental results of secondary convolution during the read-out stage. Panel (a) shows the secondary convolution of images retrieved from memory using identical control light. Panels (b) and (c) are retrieved patterns and the efficiency associated with higher-order SPC processing, respectively.}
	\label{Fig:writeandreadbothvortex}
\end{figure}

\emph{Secondary processing in the read-out stage. — } Subsequently, we further investigate the convolution process in the read-out stage. As outlined in Eq.~\ref{eqn:retrievedfield}, control lights perform an inverse convolution, transferring the conjugate spiral phase to the signal light during the read-out stage as compared to the write-in stage. In a scenario where the write-in and read-out stage topological charges are equal (i.e., $\mathcal{U}_l^W=\mathcal{U}_{l'}^R$), the read-out convolution counteracts the edge enhancement achieved during the write-in stage. FIG.~\ref{Fig:writeandreadbothvortex} (a) presents the results for this scenario when $l=l^{\prime} \in [1,6]$. The results, which show strong agreement between theoretical predictions and experimental observations, indicate that the secondary convolution nullifies the edge enhancement applied to stored images during the read-out stage. As the values of $l$ and $l^{\prime}$ increases, a dark-field effect emerges in the retrieved images, characterized by enhanced contrast at the edges. This effect arises because the high-frequency cutoff of the HyGG mode, which propagates through a finite aperture optical system (the experiment used one-inch lenses), increases with the topological charge. Consequently, this results in a hollow control light in the optical interaction region; further details can be found in the supplemental material.

The secondary convolution facilitates higher-order SPC processing in the retrieved images when considering more general scenarios, where  $\mathcal{U}_l^W\ne \mathcal{U}_{l'}^R$. The effect of higher-order SPC changes with the difference in topological charges, denoted as  $\Delta l=\left|l-l^{\prime}\right|$. Two examples of pattern ‘$\bm{\Psi}$’ for cases where  $\Delta l$=2 and 3 are depicted in Figure 3(b). Observations using the ICCD agree well with the predictions from simulations.
%%All of the results validate the protocol of all-optical convolution within the PIM framework based on EIT-based optical memory. 
Another aspect worth investigating is the transmission efficiency of signal light in this PIM-based all-optical analog computation. The likelihood of successfully writing and retrieving photons from the memory is highly contingent on the spatial overlap between the interacting waves within the ensemble. Specifically, control light that facilitates phase-only convolution, without amplitude modulation, can offer higher and consistent efficiency, regardless of the computational content.
In this proof-of-principle experiment, however, we expect that the aforementioned high-frequency cutoff (or the size of the hollow area) of the control light increases with the topological charge it carries, leading to decreased efficiency. To examine this, we stored a signal mode using two methods: (i) maintaining identical topological charges for control lights during both the write-in and read-out stages (i.e.,  $\mathcal{U}_l^W=\mathcal{U}_{l'}^R$), and (ii) utilizing a flat wavefront control light during the read-out stage (i.e.,$U_{l'=0}^R$). The normalized integrated photon counts recorded by the ICCD at $\tau_s=50$ $\mu$s are shown in FIG.~\ref{Fig:writeandreadbothvortex} (c). A donut-shaped retrieval pattern is observed when the signal field is in the fundamental Gaussian mode, along with a more pronounced decrease in efficiency for approach (i), confirming our inference. 
Nonetheless, the retrieval efficiency at $\left|l\right|=3$ in both approaches remains above $1/e$ of that observed without SPC processing. By utilizing higher-aperture lenses and topological-charge-independent vortex light, such as perfect \cite{POVOL} or flat-top vortices  \cite{ZhuzhihanflattopOAM,FlattopOL}, the unwanted amplitude modulation and its subsequent impact on efficiency can be effectively mitigated.

\section{Conclusion}
We conducted an experimental investigation into all-optical convolution using p rocessing-in-memory based on a cold atomic ensemble. The apparatus was set up with a 4f optical convolution configuration, placing the atomic ensemble at the spectral plane. This ensemble functions as a computational storage device, capable of performing image convolution, storing the results, and releasing them upon request for further processing. 
By using phase transfer between signal and control fields within the EIT protocol, we demonstrated all-optical convolution of control fields with the Fourier spectrum of images during both the write-in and read-out stages. The edge recognition of the retrieved signal field aligned closely with theoretical predictions. 
Furthermore, we extended the available duration for computation and storage to over 320 $\mu s$. These results substantiate the feasibility of developing computational storage with cold atoms and promise a wide range of applications in future all-optical processing studies.

\section{ackonwledgements}
We acknowledge the fundings from National Natural Science Foundation of China (Grant Nos. T2495253, 12304561, 12474324, U20A20218, 61525504, and 61435011), National Key R$\&$D Program of China (Grant No. 2022YFA1404002),  Anhui Initiative in Quantum Information Technologies (Grant No. AHY020200), Major Science and Technology Projects in Anhui Province (Grant No. 202203a13010001), and Youth Innovation Promotion Association of the Chinese Academy of Sciences (Grant No. 2018490).

\bibliography{spcwithmemory.bib}% Produces the bibliography via BibTeX.

\end{document}